\documentclass[prb,twocolumn,showpacs,superscriptaddress,preprintnumbers,amsmath,amssymb]{revtex4}
\usepackage{graphicx} \usepackage{dcolumn} \usepackage{bm} \usepackage{amsmath} \usepackage{color}
\usepackage{hyperref} \usepackage[utf8]{inputenc} \usepackage[T1]{fontenc}

\begin{document}


\title{Enhanced quasiparticle dynamics of quantum well states: the giant Rashba system BiTeI and topological insulators}


\author{V. Gnezdilov} \affiliation{Institute for Condensed Matter Physics, Technical University of Braunschweig, Mendelssohnstr. 3, D-38106 Braunschweig, Germany}
\affiliation{B. I. Verkin Institute for Low Temperature Physics and Engineering of the National Academy of Sciences of Ukraine, Kharkov 61103, Ukraine}

\author{P. Lemmens} \affiliation{Institute for Condensed Matter Physics, Technical University of Braunschweig, Mendelssohnstr. 3, D-38106 Braunschweig, Germany}

\author{D. Wulferding} \affiliation{Institute for Condensed Matter Physics, Technical University of Braunschweig, Mendelssohnstr. 3, D-38106 Braunschweig, Germany}

\author{A. M\"{o}ller} \affiliation{Department of Chemistry and Texas Center for Superconductivity, University Houston, Houston, Texas 77204-5003, United States}

\author{P. Recher} \affiliation{Institute for Mathematical Physics, Technical University of Braunschweig, Mendelssohnstr. 3, D-38106 Braunschweig, Germany}

\author{H. Berger} \affiliation{Institute de Physique de la Matiere Complexe, EPFL, CH-1015 Lausanne, Switzerland}

\author{R. Sankar} \affiliation{Center for Condensed Matter Sciences, National Taiwan University, Taipei, 10617 Taiwan}

\author{F. C. Chou} \affiliation{Center for Condensed Matter Sciences, National Taiwan University, Taipei, 10617 Taiwan}


\date{\today}

\begin{abstract}
In the giant Rashba semiconductor BiTeI electronic surface scattering with Lorentzian linewidth is observed that shows a strong dependence on surface termination and surface potential shifts. A comparison with the topological insulator Bi$_2$Se$_3$ evidences that surface confined quantum well states are the origin of these processes. We notice an enhanced quasiparticle dynamics of these states with scattering rates that are comparable to polaronic systems in the collision dominated regime. The E$_g$ symmetry of the Lorentzian scattering contribution is different from the chiral (RL) symmetry of the corresponding signal in the topological insulator although both systems have spin-split surface states.

\end{abstract}


\maketitle

\section{Introduction}

In the developing field of spintronics, materials are investigated that allow to manipulate and use the electron spin for information processing, e.g. by taking advantage of spin induced energy splittings of electronic states. Only recently in BiTeI a so called giant Rashba effect has been discovered with a spin splitting of the order of 0.4 eV at the Fermi energy.~\cite{Ishizaka11} Rashba coupling leads to a spin polarization depending momentum shift $k_R$ and a resulting spin splitting of the electronic dispersions.~\cite{Bahramy12,rashba60} Up to now spin splittings of the order of several meV in zero magnetic fields have been found in InGaAs/InAlAs heterostructures,~\cite{Das89} semiconductor interfaces,~\cite{Nitta97} and non-magnetic metallic surfaces.~\cite{Lashell96, Koroteev04, Ast07} For silver surfaces covered by Bi a splitting of the order of 200 meV has been reported.~\cite{Ast07}

BiTeI is a layered, polar semiconductor and crystallizes in the primitive trigonal space group, $P3m1$ = $C^1_{3v}$, No. 156, with one formula unit per unit cell (Z=1). Its crystal structure exhibits triangular layers of the respective chemical elements with a fixed stacking sequence along the crystallographic \emph{c}-axis, see Fig. 1. The chemical bonding between Bi and Te can be described as more covalent in character than the Bi-I interaction. Nevertheless, both anions are involved in covalent bonding with Bi and van-der-Waals type interactions across the void layers ($\Box$). The latter is a much weaker interaction and thus allows two different surfaces, Te- and I-terminated, to be obtained by cleaving a single crystal to a ``top'' and ``bottom'' piece. This situation leads to a sufficient electric gradient that can significantly enhance Rashba coupling.~\cite{Bell13}

The Rashba induced spin splitting in BiTeI is due to the combination of three effects:~\cite{Bahramy11,Eremeev12b,Zhu13} a large atomic spin-orbit interaction in an inversion-asymmetric media, a narrow band gap, and the same symmetry of highest energy valence and lowest energy conduction bands. Photoemission,~\cite{Ishizaka11} optical absorption,~\cite{Lee11} and Shubnikov-de Haas oscillations~\cite{Bell13} have been used to characterize the Rashba-split electronic branches. In Raman scattering a general softening of phonon frequencies in BiTeI and BiTeCl has been attributed to spin-orbit coupling.~\cite{Sklyadneva12} There even exist evidence for a transition into a topologically ordered state under high pressure.~\cite{Bahramy12,Tran13,Xi13}.

BiTeI also shows Rashba-split surface states.~\cite{Ishizaka11,Landolt12,Crepaldi12} Interestingly these states are ambipolar on differently terminated surfaces. On Te-terminated surfaces electron like, occupied surface states are induced. Charge accumulation, a resulting down shift of the conduction band and the confinement of these states into quantum well states evolve as function of time as shown by ARPES experiments. In contrast, their spin splitting shows no dependence on the surface potential.~\cite{Landolt12,Crepaldi12}. A similar evolution with time is observed in GaInN/GaN heterostructures, where charging effects lead to the observation of a time dependent intensity of spontaneous light emission.~\cite{Thomsen11} The Rashba splitting of the surface states as well as other properties have been carefully theoretically studied.~\cite{Eremeev12b,Eremeev12} However, to our knowledge there exist no further experiments beyond ARPES to enlighten their properties.

The topological insulators, e.g. Bi$_2$Se$_3$, are interesting counter parts to the giant Rashba systems. They are also small gap ($\Delta = 0.3$ eV), degenerate semiconductors and share a layered, rhombohedral crystal structure ($D^5_{3d}$). For the two compounds the interatomic distances are very similar and the size of the gap in Bi$_2$Se$_3$ and the spin splitting in BiTeI are of similar magnitude. In Bi$_2$Se$_3$, however, bulk inversion symmetry leads to a different role of spin-orbit coupling. Here, only the surface shows Dirac states with topological order while the bulk states are spin degenerate.~\cite{Zhang09,Kane11,Hasan11} Inelastic scattering processes of the Dirac states to bulk states have recently been discussed for scanning tunneling spectroscopy~\cite{Kim11} as well as Raman experiments. These processes might be related to the limitation of topological protection to elastic backscattering.~\cite{Gnezdilov11}

Bi$_2$Se$_3$ surfaces also show time evolutions of their electronic structure. In photoemission experiments time dependent shifts of conduction bands and the Dirac states below $E_F$ have been observed.\cite{Zhu11}  These shifts are induced by charge accumulation and band bending with characteristic time constants in the range of 20 - 40 hours. Surface reactions with water or K and Rb doping have been used to study these effects.~\cite{Benia11} The charge confinement in a surface layer leads to quantum well states that are also Rashba split.~\cite{Benia13} It is therefore of interest to compare the giant Rashba system BiTeI with topological insulators to achieve a better understanding of the surface scattering processes and their evolution with charge accumulation. This comparison could shed some light on the role and interplay of spin-orbit coupling with other degrees of freedom in these materials.

Quantum well states in semiconductor heterostructures have previously been intensively studied using Raman scattering.~\cite{Pinczuk89,Prevot91} Collective charge density as well as single particle spin excitations show up as maxima in parallel and crossed polarization, respectively. Resonances due to intersubband excitations or over the bulk optical gap E$_0$+$\Delta$$_0$ allow a sufficiently large light-matter coupling and enhance the sensitivity of the Raman scattering experiment.\cite{Pinczuk89} Effects of quantum confinement, free carrier generation by band bending, and photodoping have been investigated with carrier concentrations as low as n<10$^{11}$/cm$^2$.

We have performed Raman scattering experiments on BiTeI single crystals showing bulk phonon and electronic excitations with a pronounced resonance as a function of the wavelength of the incident laser radiation. Low energy excitations that are attributed to surface scattering are observed only on one type of surface. This agrees very well with band structure calculations of the electronic properties of the Te terminated surfaces that shows occupied electron like states with band bending.~\cite{Eremeev12b,Eremeev12} From our observations we follow an enhanced quasiparticle dynamics of the quantum well states. Despite differences in the electronic scattering rates and the symmetry of the signal there is a surprising similarity of the observed features with the topological insulator Bi$_2$Se$_3$.

\section{Experimental}

Raman scattering experiments were performed in quasi-backscattering geometry using a $\lambda = 532$ nm solid state laser and single crystals prepared both by transport and Bridgman techniques. We did not observe significant differences between samples originating from the two preparation techniques. Thus we will only show data from samples prepared by Bridgman growth. Structural and transport experiments have been performed as a characterization.~\cite{Wang13} The used batch shows a specific resistivity $\rho$=3.2$\pm 0.36$ (m$\Omega$cm), an RRR(300K/2K)=2, a charge density n(5K)=-21$\pm1.9$ ($\times$10$^{18}cm^{-3}$), and mobility $\mu$(2K)=180$\pm19$ (cm$^2$/Vs).

Circular light polarizations are denoted by $RR$ and $RL$, with right-handed and left-handed circular polarized incident and scattered light. To probe resonances of the scattering cross sections we used different laser lines of an argon-krypton mixed ion laser. The laser power was set to 5 mW with a spot diameter of approximately 100 $\mu$m to avoid heating effects and sample deterioration. The melting point of BiTeI is only 520$^\circ$C. All measurements were carried out in an evacuated closed cycle cryostat in the temperature range from 6.5 K to 295 K. The spectra were collected using a triple Raman spectrometer (Dilor-XY-500) with an attached liquid nitrogen cooled CCD (Horiba Jobin-Yvon, Spectrum One CCD-3000V).

Freshly cleaved sample surfaces were prepared at ambient conditions using scotch tape from the ``top'' as well as from the ``bottom'' of large single crystals (typical size $4\times4\times2~mm^3$). Cleaved-off pieces as well as their opposite faces are rapidly cooled down in vacuum to minimize surface degradation. This preparation leads to two surface terminations, \textit{Te} and \textit{I}, see Fig. 1.



\section{Results and Discussion}
\subsection{Phonon scattering}

\begin{figure}
\label{figure1}
\centering
\includegraphics[width=8cm]{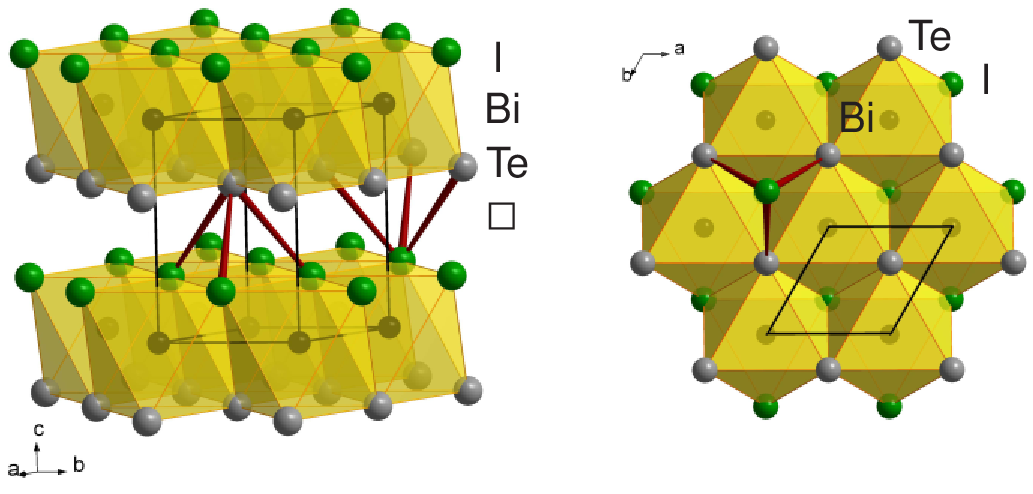}
\caption{(Colour online) Sketch of the crystal structure of BiTeI in two projections. Layers of I, Bi, Te, and voids ($\Box$) are marked. Selected bonds that are cut by cleaving and establish the Te and I terminated surface are drawn by thick lines. }
\end{figure}

Low-frequency Raman spectra of BiTeI from the $ab$-plane in parallel ($XX$), crossed ($YX$), and two circular polarization configurations are shown in Fig. 2. The spectra are well polarized and four strong lines can be easily identified in the spectra. According to the space group $P3m1$, the Bi, Te, and I have site symmetries given by \emph{1c}, \emph{1b}, and \emph{1a}, respectively. This leads to $\Gamma = 2A_1 + 2E = 4$ Raman-active phonon modes. Lines located at 93 and 150 cm$^{-1}$ are assigned to $A_1$(1) and $A_1$(2) phonon modes. The lines at 55 and 102 cm$^{-1}$ correspond to $E$(1) and $E$(2) phonon modes, respectively. These phonon lines are superimposed by a structured but weaker background of probably defect or electronic scattering origin (see Fig. 3). With exception of the $A_1$(2) mode at 150 cm$^{-1}$ all Raman active phonon have a symmetric line shape. We interpret the asymmetric line shape of the latter mode as due to coupling to electronic degrees of freedom (Fano line shape). A similar asymmetry has been observed in optical absorption of a mode at 143 cm$^{-1}$.~\cite{Martin12} The spectra from samples grown by Bridgman and transport techniques are identical with respect to the phonon frequencies. Slight deviations exist with respect to the intensity of some phonons and the background scattering. The shown data are in general agreement with an earlier Raman scattering and band structure investigation of BiTeI and BiTeCl.~\cite{Sklyadneva12}

\begin{figure}
\label{figure2}
\centering
\includegraphics[width=8cm]{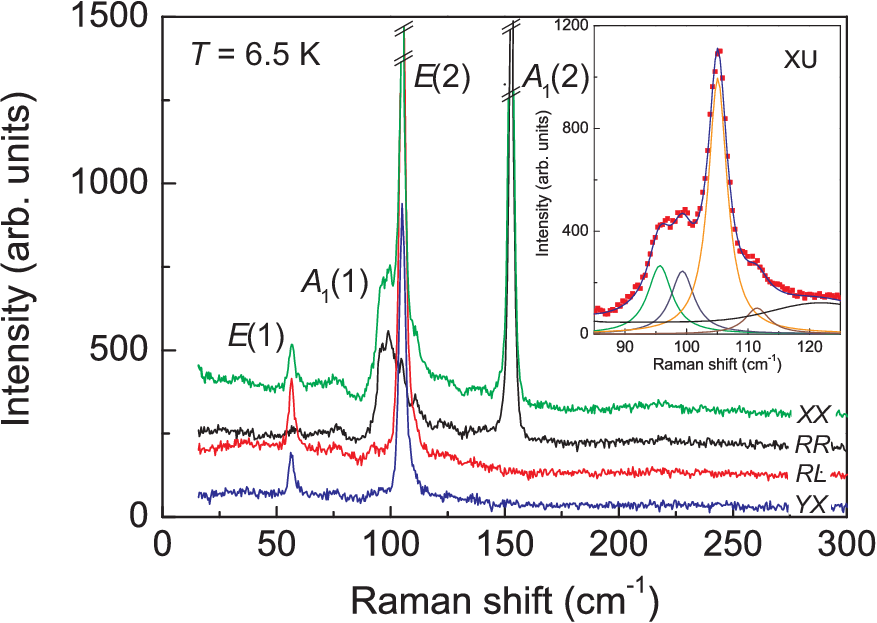}
\caption{(Colour online) Polarized Raman spectra in different scattering configurations of the single crystal surface, including circular polarizations of BiTeI at $T=6.5$ K and $\lambda = 532$ nm. Spectra are shifted in intensity for clarity. The phonons with higher scattering intensity are cut off to emphasis signals with smaller intensity. The inset gives a fit by Lorentzian lines to an unpolarized Raman spectrum.}
\end{figure}

The low temperature data with generally small linewidths allow a fitting of the phonons to individual Lorentzian lines, shown in Fig. 2. According to this analysis and general knowledge on light scattering in semiconductors ~\cite{Loudon64} there exists a splitting of the dipole-active lattice vibrations of the order of 5 cm$^{-1}$ into doublets of longitudinal (LO) and transverse (TO) vibrations. These doublets are observed as in our experiment the scattering wave vector \textbf{\textit{q}} has a component perpendicular to the crystallographic $c$ axis. We have omitted a further analysis of these modes as these aspects are not in the center of our investigation and refer to Ref.\cite{Sklyadneva12}.

Raman scattering with different incident Laser energies may lead to important information about electronic states involved in the scattering process. Respective normalized phonon intensities with laser energies in the energy range 1.96 to 2.54 eV (632nm to 488nm) at T=6.5 K are shown in Fig. 3. They show a large change in the high photon energy regime with the intensities of three phonon modes behaving rather similar. We follow a resonant enhancement of the intensity at $\sim 2.4$ eV. This characteristic energy fits to the electronic dispersions of BiTeI at the $\Gamma$ point as there exist several interband transitions with this separation.~\cite{Bahramy11, Ishizaka11} Therefore, we attribute the observed resonance in Raman scattering to an interband transition. Also in the topological insulator Bi$_2$Se$_3$ a resonant enhancement of the scattering intensity has been observed. At low temperatures it leads to a sufficient amplification to observe surface induced scattering.~\cite{Gnezdilov11}

\begin{figure}
\label{figure5}
\centering
\includegraphics[width=8cm]{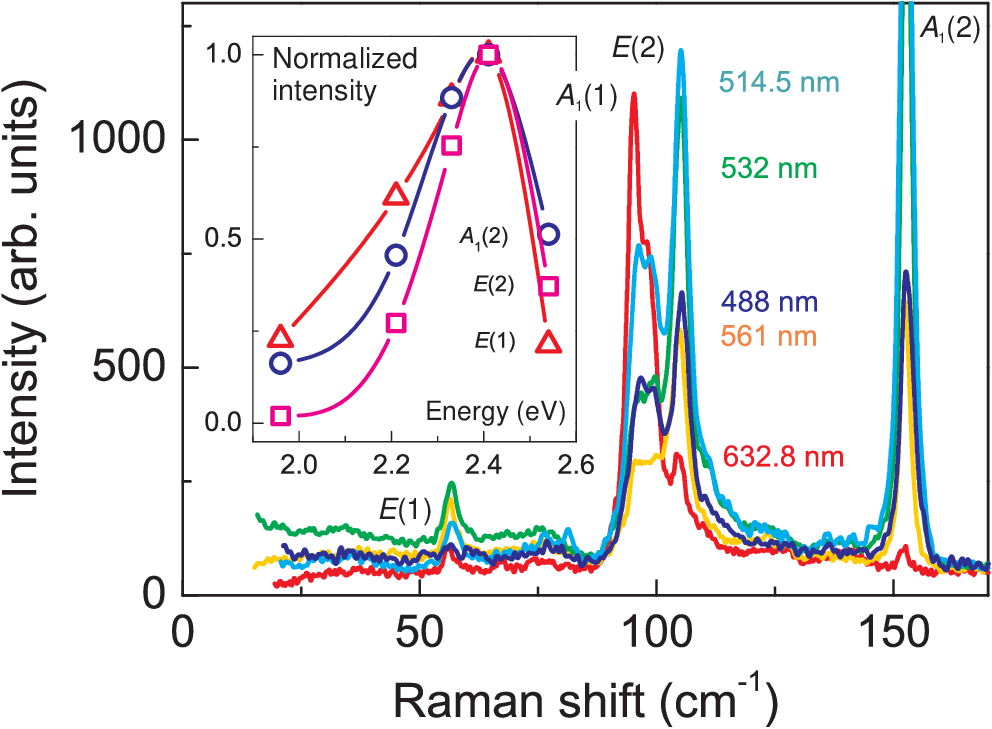}
\caption{(Colour online) Raman spectra of BiTeI with different incident Laser lines. Experiments are performed in \emph{XX} polarization and at T=6K. The inset gives the normalized intensity for three phonon lines as function of incident Laser energy (eV).}
\end{figure}

\subsection{Electronic Raman scattering}

\begin{figure}
\label{figure7}
\centering
\includegraphics[width=8cm]{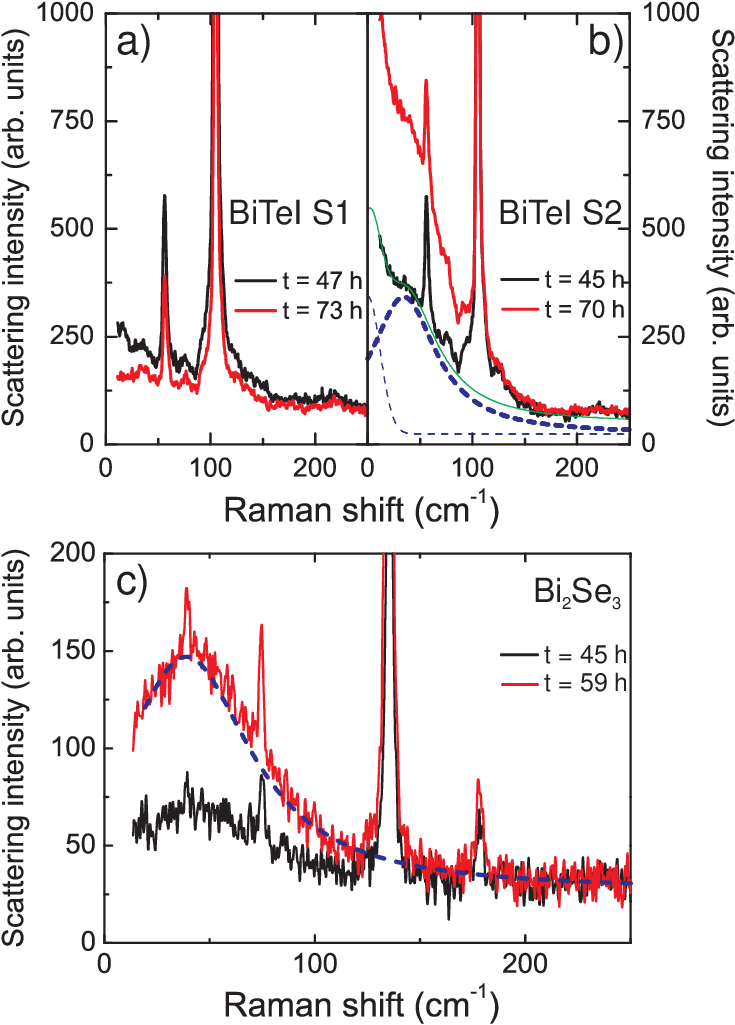}
\caption{(Colour online) Raman spectra emphasizing the low energy regime a) of BiTeI obtained on Surface 1 (S1) and b) obtained on Surface 2 (S2), and c) of Bi$_2$Se$_3$. The spectra were measured approximately two (black curve) and three (red curve) days after cleaving the single crystal surface ($T = 6.5$ K and $RL$ polarization). The dashed lines gives fits to the low energy continuum using a Lorentzian and a quasi-elastic line. }
\end{figure}

In the following we will discuss effects that we attribute to electronic surface states of BiTeI. This assignment is based on their spectral range and phenomenology as discussed below. Sufficient sensitivity to detect surface scattering processes exist due to the resonant enhancement of the scattering intensity demonstrated in Fig. 3.

Fig. 4 a) and b) shows Raman spectra from two differently terminated surfaces denoted by S1 and S2. They correspond to two freshly prepared, opposite faces obtained by cleaving a single crystal. With respect to phonon scattering there is no difference observed in the two surfaces. This is due to their origin in bulk excitations. In contrast, S2 shows an additional low energy signal for frequencies below $\Delta\omega\approx 150$ cm$^{-1}$, indicated by the solid green line in Fig. 4 b. The linewidth of this signal is much broader than the previously discussed phonon scattering and it is repeatably observed after consecutive cleaving steps along the crystal's layer stacking. The difference between the two surface terminations S1 and S2 evidences that stacking faults are not observed here and that the surface termination is not detoriated by cleaving.

The low energy scattering (dashed lines) is essentially given ba a Lorentzian (linewidth $w = 75$ cm$^{-1}$, energy $E_{max} = 34$ cm$^{-1}$) and quasi-elastic scattering ($E\approx 0$). With respect to selection rules, the Lorentzian has dominant intensity in \textit{XX} and \textit{RL} polarization, similar to the E bulk phonons. Therefore we assign it to E symmetry. The low energy scattering observed on S1 is much weaker in intensity. This scattering could also be a remnant of the weakly temperature dependent continuum that we described earlier and attributed to defects. In Fig. 4 c) we compare these results with the topological insulator Bi$_2$Se$_3$. Table 1 gives further cases of low energy, Lorentzian scattering with fit parameters.

\begin{table}
\caption{Maxima and line width of collision dominated Raman scattering with Lorentzian linewidth in several compounds. In Eu$_{1-x}$Gd$_{x}$O the maximum position and the scattering rate is a function of temperature. }
\begin{tabular}{|c|c|c|c|}
  \hline
  Compound        &$E_{max}$     & Width (FWHM) & Reference\\
                  & [cm$^{-1}$]  & [cm$^{-1}$]  & \\ \hline
  BiTeI           & 34           & 75        & this work \\
  Bi$_2$Se$_3$    & 39           & 80         & \cite{Gnezdilov11} \\
  Eu$_{1-x}$Gd$_{x}$O & 20-45, f(T)    & -         & \cite{Rho02} \\
  Na$_x$CoO$_2$, x=0.78& 58             & -         & \cite{Lemmens06} \\
  \hline
\end{tabular}
\end{table}

The low energy scattering shows a remarkable time dependence of its intensity without shifting in energy, see two subsequent measurements given in Fig. 4b) and a detailed analysis of the integrated intensity in Fig. 5. The intensity starts to increase about 12 h after the cleavage and increases further until it reaches a maximum after about 70h. For very long times, at about 100 h after cleavage, the quasi-elastic tail dominates the low energy regime and it becomes difficult to separate it from the Lorentzian. Layered halides and chalcogenides are prone to non-stoichiometry and surface modifications. Still, the weak dependence of the lineshape on time and their reproducibility with repeated cleaving gives evidence for intrinsic physics involved. Furthermore, the observed time dependence resembles charging effects in GaInN/GaN quantum well structures ~\cite{Thomsen11} and on Bi$_2$Se$_3$ surfaces \cite{Zhu11}. In the inset of Fig. 5 a Raman spectrum of Bi$_2$Se$_3$ at T=3K in \emph{RL} polarization is shown. The Lorentzian shows a very similar evolution with time with similar line width and energy, see Table 1. Also photo emission experiments found time dependent shifts of the chemical surface potential due to reactive doping by $H_2O$ or K.\cite{Benia11} We assume a very similar surface chemistry that is induced by cleaving at ambient conditions. This effect may be used to monitor the electronic properties of the surface states.

Low energy, Lorentzian Raman scattering described by $ S(q,\omega) \propto (1+n(\omega))\cdot(\omega \Gamma)/(\omega^{2}+ \Gamma^{2})$ is usually attributed to collision dominated electronic processes.~\cite{varma89,zawadowski90} Here, $\Gamma$ is the scattering rate that might itself be frequency dependent and $n$ is the Bose Factor. Collision dominated processes prevent the usual screening of electronic intraband excitations.~\cite{Rho02,Lemmens06} A well defined Lorentzian maximum is observed with $E_{max} \approx \Gamma $ if there exists a single, dominant scattering rate. For $\Gamma \approx 1/\omega$ a plateau of scattering is observed as exemplified by high temperature superconductors.\cite{Devereaux} Table 1 gives an overview of peak positions in $Eu_{1-x}Gd_{x}O$ that is in the proximity to a metal insulator transition~\cite{Rho02}, $Na_xCoO_2$ with spin polarons and the topological insulator Bi$_2$Se$_3$. The peak position of BiTeI is very well comparable to Bi$_2$Se$_3$. Respective Raman data are shown in Fig. 4c) for comparison.

\begin{figure}
\label{figure7}
\centering
\includegraphics[width=8cm]{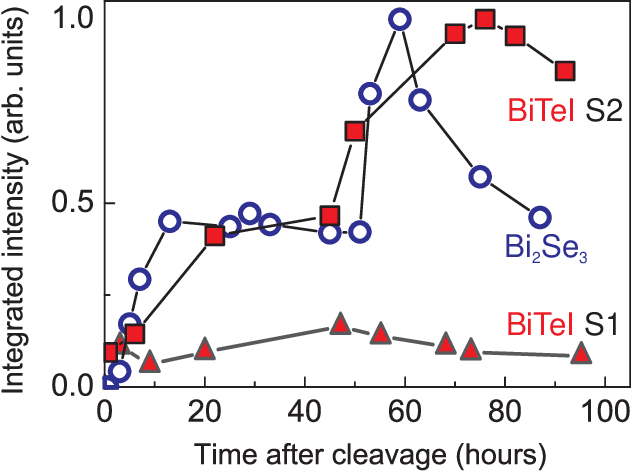}
\caption{(Colour online) Time evolution of the low energy scattering intensity for both surfaces, S1 and S2 as full triangles and full squares, respectively. The time dependence of a similar signal in the topological insulator Bi$_2$Se$_3$ is shown with open circles. }
\end{figure}

In Bi$_2$Se$_3$ the Lorentzian Raman scattering maximum has been attributed to surfaces states based on its chiral, i.e. \textit{RL}, symmetry.~\cite{Gnezdilov11} In the inversion symmetric Bi$_2$Se$_3$ such a symmetry component is not allowed in the bulk of the crystal. Based on this symmetry it has also been argued that the signal must involve inelastic scattering processes that connect the topological protected Dirac states with continuum states.\cite{Kim11} Furthermore, also comparable time dependencies of the scattering intensity have been observed that kept peak energy and linewidth invariant, similar to the data of BiTeI shown in Fig. 5.

At this point we contrast our observations with light scattering experiments in modulation doped (Ga,Al)As multi-quantum wells.~\cite{Prevot91} In crossed polarization such systems show spin flip scattering due to an intersubband excitation with a maximum at 174 cm$^{-1}$. This energy is identical with the lowest intersubband excitation and changes with the binding potential, i.e. it depends on sample properties. In parallel polarization coupled LO phonon to collective intersubband charge excitation are observed at 224cm$^{-1}$ and 310cm$^{-1}$, respectively. Transferring this to the case of BiTeI we would expect modes in the energy range $\Delta$$_{B1-B2}$$\approx$120meV=960cm$^{-1}$, $\Delta$$_{B2-B3}$$\approx$50meV=400cm$^{-1}$ for intersubband excitations in Bi$_2$Se$_3$ and similar energies in BiTeI.~\cite{Benia11,Crepaldi12} For the latter binding energies have not been resolved in ARPES so clearly.\cite{Crepaldi12}. Nevertheless, the discrepancy in energy and its missing time dependence are clear arguments against an interpretation as intersubband excitations. Furthermore, our experiments show the Lorentzian modes in E symmetry and in RL symmetry for Bi$_2$Se$_3$, respectively. There is also no evidence for coupled LO phonon charge excitations. All this supports our assignment of the Lorentzian maxima to collision dominated scattering.

The concerted behavior of the two chalcogenides calls for their joint description. One key seems to be the time variations of intensity. Photoemission shows a very similar time evolution of the confinement of charge carriers into quantum well states. For both compounds these surface states are Rashba split and have a chiral spin texture. For BiTeI the involvement of quantum well states explains the selective observation of Lorentzian Raman on one of the two surface terminations. The observation of an E symmetry of the scattering as well as the additional quasi-elastic signal differ from observations in Bi$_2$Se$_3$. This could be related to details of the electronic structure of the two compounds. Nevertheless, a scenario based on Rashba spin split quantum well states does not need to take into account the spin momentum locking of the Dirac states.

Band structure calculations for BiTeI show that the enormous Rashba spin splitting leads to two differently spin polarized Fermi surfaces, with a pronounced hexagonal warping of the higher energy sheet. It has been found that larger in-plane potential gradients, corresponding to the localization of states, lead to a larger out-of-plane $S_z$ component.~\cite{Eremeev12b} This component in conjunction with warping is in our opinion responsible for collision dominated scattering similar to polaronic systems. Warping and nesting of Fermi surfaces generally lead to an anisotropic selection of scattering vectors enhancing scattering~\cite{Virosztek92} and enhancing the related Raman scattering intensity.~\cite{Devereaux} To our knowledge this is the first example of a collision dominated regime of quantum well states. In contrast, in GaAs heterostructures they are related to high electronic mobilities. An enhanced quasiparticle dynamics for BiTeI surface states has also been proposed by theory. Rough estimations of the corresponding scattering rate show a reasonable agreement with our experimental data, see Fig. 3d) in Ref.~\cite{Eremeev12b}.

With respect to the topological insulator Bi$_2$Se$_3$ all previous arguments hold, e.g. there exist a warped surface with pronounced $S_z$ components~\cite{Hasan11} if the Dirac point is shifted far enough below E$_F$. The chiral symmetry of collision dominated scattering with one well defined scattering rate for Bi$_2$Se$_3$ suggests that the related states have only very little overlap with bulk states or that the scattering is very anisotropic as demonstrated in some STM experiments.~\cite{Kim11,alpichshev10} A recent ARPES experiment has shown that cleaving single crystals at ambient conditions lowers the Dirac cone even further, enhances the Fermi surface warping, while keeping the topological protection intact.~\cite{chen13} This explains the strong dependence of the electronic Raman scattering on time.

The low dimensionality of quantum well states suggest them to be candidates for case studies on the effects of 3D to 2D crossover or transitions between trivial and topological phases.\cite{Qi11} Such a transition may be induced by an increasing hybridization between surface and bulk states, i.e. by scattering processes that are also relevant for the observed collision dominated regime of the quantum well states. We should, however, be reminded that these states themselves are not topologically protected. This could lead to interesting anomalies with increasing but still weak disorder.\cite{Kettemann} To our knowledge, quantum well states induced by larger surface potential shifts have so far not been considered or studied in transport experiments on topological insulators. Experiments on surface doped topological insulators and giant Rashba systems could open up a promising field for future research.

\section{Summary}
In summary, Raman scattering experiments in BiTeI show a pronounced resonance of bulk phonons and electronic scattering. Low energy excitations are observed that decisively depend on surface termination. The underlying states show a high sensitivity on charging and band bending effects exemplified by a dependence of the scattering intensity on the time after cleavage at ambient condictions. The Lorentzian contribution to this scattering is modeled as collision dominated scattering with a scattering rate of $\Gamma = 34$ cm$^{-1}$ (1.02THz). The observation of these signals with comparable scattering rates in both BiTeI and the topological insulator Bi$_2$Se$_3$ point to their joint origin in quantum well states with an enhanced quasiparticle dynamics. The latter is most probably related to warping and the anisotropic selection of scattering processes in conjunction with a large $Sz$ component. The scattering rates are therefore comparable to those of polaronic systems. In the topological insulator these signals exhibit a chiral symmetry that is forbidden in the bulk of an inversion symmetric system. In BiTeI this scattering is of $E_g$ symmetry, an allowed symmetry in the bulk. Therefore, we propose further transport experiments for both classes of materials to investigate the quantum well states. This could further elucidate the related quasiparticle dynamics and search for possible disorder or hybridization induced topological phase transitions.

\begin{acknowledgments}
We acknowledge important discussions with Ch. Ast, S. Eremeev, and S. Kettemann. We thank K. Schnettler for important help and acknowledge support by DFG, B-IGSM and the NTH School \textit{Contacts in Nanosystems}.
\end{acknowledgments}

\end{document}